\documentclass[twocolumn,letter]{jpsj3}
\usepackage{color}
\usepackage{braket}
\usepackage{mathrsfs}

\usepackage{graphicx}
\usepackage{graphics}
\usepackage{amssymb, latexsym}
\usepackage{amsmath}
\usepackage{bm}
\usepackage{color}
\usepackage{ulem}
\usepackage{fancybox}
\usepackage{xspace}
\def\runauthor{Y. Akagi, M. Udagawa, and Y. Motome}

\title{Effect of Quantum Spin Fluctuation on Scalar Chiral Ordering \\
in the Kondo Lattice Model on a Triangular Lattice
}

\author{Yutaka Akagi\thanks{E-mail address: akagi@aion.t.u-tokyo.ac.jp}, Masafumi Udagawa, and Yukitoshi Motome 
}
\inst{Department of Applied Physics, University of Tokyo, 
Hongo 7-3-1, Bunkyo-ku, Tokyo 113-8656
}
\abst{Spin scalar chiral ordering gives rise to nontrivial topological characters and peculiar transport properties.  
We here examine how quantum spin fluctuations affect the spin scalar chiral ordering in itinerant electron systems. 
We take the Kondo lattice model on a triangular lattice, and perform the linear spin wave analysis 
in the Chern insulator phases with spin scalar chiral ordering obtained in the case that the localized spins are classical. 
We find that, although the quantum fluctuation destabilizes the spin scalar chiral phase 
at 3/4 filling that originates from the perfect nesting of Fermi surface, 
it retains the phase at 1/4 filling that is induced by an effective positive biquadratic interaction.
The reduction of the ordered magnetic moment by the zero-point quantum fluctuation is considerably small, 
compared with those in spin-only systems. The results suggest that the Chern insulator at 1/4 filling remains robust 
under quantum fluctuations.
}

\kword{spin scalar chirality, spin wave theory, Kondo lattice model, frustration, triangular lattice}

\begin{document}
\maketitle
Geometrically frustrated magnets often exhibit unconventional orders and fluctuations because of the competing interactions. 
Such unconventional magnetism has been recently drawing considerable attention by the nontrivial magnetic and electronic responses. 
A representative example is the multiferroics: e.g., the coexistence of ferroelectricity and a peculiar spiral magnetic order~\cite{Kimura2003,Katsura2005,Sergienko2006}. 
Another interesting example is a noncoplanar spin texture and associated anomalous transport in itinerant magnets~\cite{Loss1992,Ye1999,Ohgushi2000,Tatara2002}. 
There, the spin scalar chirality, characterized by a non-vanishing value of the triple product of spins,
$\bm{S}_1 \cdot (\bm{S}_2 \times \bm{S}_3)$,
gives rise to internal magnetic flux for itinerant electrons through the spin Berry phase mechanism, 
leading to the novel transport, termed the topological Hall effect. 
For instance, a noncoplanar topological spin texture in ferromagnets called skyrmion has recently attracted much attention for potential applications in spintronics~\cite{Muhlbauer2009,Yu2010,Yu2012}.

\begin{figure}[t]
\begin{center}
\includegraphics[width=7.5cm]{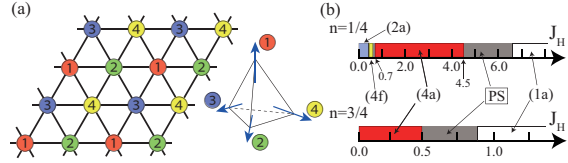}
\end{center}
\caption{(Color online) (a) Schematic picture of the four-sublattice all-out chiral order on a triangular lattice. 
The arrows on a tetrahedron show the spin directions of the corresponding sites on the triangular lattice.
(b) Phase diagram of the model in Eq.~(\ref{Ham}) obtained by the variational calculation~\cite{Akagi2010}. 
The upper (lower) figure corresponds to the phase diagram at $n=1/4$ ($n=3/4$).
(1a), (2a), (4a), and (4f) indicate the following orders:
(1a) a ferromagnetic order, (2a) a two-sublattice collinear stripe order, (4a) the four-sublattice all-out order which corresponds to the one in (a), 
and (4f) a four-sublattice coplanar $90^{\circ }$ spiral order.  
PS indicates a phase-separated region. See also Fig.~1 in Ref.~\citen{Akagi2010}. 
}
\label{chiral-order_with-PD}
\end{figure}

Recently, a realization of such a spin scalar chiral order was discovered in a minimal theoretical model, 
the Kondo lattice model with classical localized spins on a triangular lattice [see Eq.~(\ref{Ham})]. 
The model was first shown to exhibit a four-sublattice noncoplanar magnetic order at 3/4 filling, 
which has ferroic ordering of the spin scalar chirality defined in each triangle unit [Fig.~\ref{chiral-order_with-PD}(a)].~\cite{Martin2008} 
This state can be viewed as a crystal of the smallest-size skyrmions.
Soon later, another spin scalar chiral phase with the same magnetic structure was shown to exist near 1/4 filling 
in a wider parameter range~\cite{Akagi2010}.
The chiral phases at 3/4 and 1/4 filling are both Chern insulators showing the quantization of the Hall conductivity. 
The stabilization mechanism of the chiral ordering, however, is different between the two phases; 
the former originates in the perfect nesting of Fermi surface~\cite{Martin2008},
whereas the latter is induced by a critical enhancement of effective positive biquadratic interactions through the generalized Kohn anomaly~\cite{Akagi2012}.

While the existence of spin scalar chiral phases is thus established for the model with classical localized spins, 
it poses a fundamental question: how does the quantum fluctuation of localized spins affect the nature of the spin scalar chiral phases? 
In general, quantum fluctuations play a significant role in frustrated magnets 
because of severe competition between many possible ordered states. 
In the present case, it is particularly interesting to compare how differently quantum fluctuations affect the two chiral phases, as they have different origins as mentioned above.

In this study, we investigate the effect of quantum fluctuations on the scalar chiral phases within the linear spin wave theory.
For this purpose, we generalize the formulation for 
the ferromagnetic state in the double-exchange model~\cite{Furukawa1996} to the four-sublattice noncoplanar magnetic order.
We show that the scalar chiral order is destabilized by quantum fluctuations at 3/4 filling, whereas it remains robust at 1/4 filling. 
We also show that the spin wave correction to the length of ordered moment is considerably small, compared with those in 
the systems with localized spins only. 

We start with the ferromagnetic Kondo lattice model on a triangular lattice, 
\begin{eqnarray}
\mathcal{H} = -t\sum\limits_{\langle i\alpha, i'\alpha'\rangle}\sum\limits_{s=\uparrow, \downarrow}\bigl(c^{\dag}_{i\alpha s}c_{i'\alpha's} + {\rm H. c.}\bigr) \notag\\
- \frac{J_{\rm H}}{S}\sum\limits_{i,\alpha}\sum\limits_{s,s'=\uparrow, \downarrow}{\mathbf S}_{i\alpha}\cdot c^{\dag}_{i\alpha s}\bm{\sigma}_{ss'}c_{i\alpha s'},
\label{Ham}
\end{eqnarray}
where $c^{\dag}_{i\alpha s}$($c_{i\alpha s}$) is a creation (annihilation) operator 
for a conduction electron with spin $s$ on site $(i,\alpha)$, 
the $\alpha$th sublattice site in the $i$th unit cell. 
Here, we consider the four-sublattice magnetic unit cell, i.e., $\alpha=1$, 2, 3, and 4 [see Fig.~\ref{chiral-order_with-PD}(a)].
\textit{t} is the transfer integral and $J_{\rm H}(>0)$ is the Hund's-rule coupling 
(the sign of $J_{\rm H}$ is irrelevant within the linear spin wave approximation),
$\bm{\sigma}_{ss'}=({\sigma}^x_{ss'},{\sigma}^y_{ss'},{\sigma}^z_{ss'})$
is a vector representation of Pauli matrices, and ${\mathbf S_{i\alpha}}$ ($|{\mathbf S_{i\alpha}}|=S$) is a localized spin on site $(i,\alpha)$. 
The sum $\langle i\alpha, i'\alpha'\rangle$ 
is taken over the nearest-neighbor sites on the triangular lattice. Hereafter, we take $t=1$ as an energy unit 
and the Planck constant divided by $2\pi$, $\hbar=1$.

We focus on the spin scalar chiral phases, and choose the spin quantization axis of itinerant electrons parallel to the 
ordered moments $\langle{\mathbf S}_{i\alpha}\rangle$ at each site. 
Setting 
$\langle{\mathbf S}_{i\alpha}\rangle = S (\sin\theta_{\alpha}\cos\phi_{\alpha}, \sin\theta_{\alpha}\sin\phi_{\alpha}, \cos\theta_{\alpha})$ 
with $\theta_1=0$, $\theta_2=\theta_3=\theta_4=\arccos(-1/3)$, $\phi_1=\phi_2=0$, $\phi_3=2\pi/3$, and $\phi_4=-2\pi/3$, 
we denote the electron spin state parallel (anti-parallel) to the localized moment 
as $|+_{\alpha}\rangle\ (|-_{\alpha}\rangle)$: 
\begin{eqnarray}
\left\{\begin{array}{ll}
|+_{\alpha}\rangle = \cos\frac{\theta_{\alpha}}{2}|\uparrow\rangle + e^{i\phi_{\alpha}}\sin\frac{\theta_{\alpha}}{2}|\downarrow\rangle\\
|-_{\alpha}\rangle =  -e^{-i\phi_{\alpha}}\sin\frac{\theta_{\alpha}}{2}|\uparrow\rangle + \cos\frac{\theta_{\alpha}}{2}|\downarrow\rangle.
\label{eq:rotated_frame}
\end{array}\right.
\end{eqnarray}
In this local frame, the Hamiltonian in Eq.~(\ref{Ham}) is written as
\begin{eqnarray}
\mathcal{H} = -t\!\!\!\sum\limits_{\langle i\alpha, i'\alpha'\rangle}\!\sum\limits_{s,s'=\pm}\!\!\bigl(\langle s_{\alpha}|s'_{\alpha'}\rangle \tilde{c}^{\dag}_{i\alpha s}\tilde{c}_{i'\alpha' s'} + {\rm H. c.}\bigr) \notag\\ 
- \frac{J_{\rm H}}{S} \sum\limits_{i,\alpha}\sum\limits_{s,s'=\pm}\tilde{\mathbf S}_{i\alpha}\cdot \tilde{c}^{\dag}_{i\alpha s}\bm{\sigma}_{ss'}\tilde{c}_{i\alpha s'},
\label{Ham2}
\end{eqnarray}
where 
$\tilde{c}_{i\alpha s}$, $\tilde{c}_{i\alpha s}^\dagger$ 
are the operators for the states in Eq.~(\ref{eq:rotated_frame}), and $\langle\tilde{\mathbf S}_{i\alpha}\rangle = S(0,0,1)$. 

We apply the spin wave approximation to the Hamiltonian in Eq.~(\ref{Ham2}) in this new frame. 
Namely, we apply the Holstein-Primakoff transformation to localized moments,
\begin{eqnarray}
\tilde{S}^+_{i\alpha} \simeq \sqrt{2S}a_{i\alpha}, \, 
\tilde{S}^-_{i\alpha} \simeq \sqrt{2S}a^{\dag}_{i\alpha}, \, 
\tilde{S}^z_{i\alpha} = S - a^{\dag}_{i\alpha}a_{i\alpha},
\label{Holstein-Primakoff}
\end{eqnarray}
by introducing the magnon creation (annihilation) operator 
$a^{\dag}_{i\alpha}$($a_{i\alpha}$) 
at site $(i,\alpha)$.
The transformed Hamiltonian can be separated into two parts, 
$
\mathcal{H} = \mathcal{H}_0 + \mathcal{H}' 
$. 
The unperturbed part $\mathcal{H}_0$ describes the interaction between itinerant electrons and static ordered moments, which is given by 
\begin{eqnarray}
\mathcal{H}_0\! =\! -t\!\!\!\sum\limits_{\langle i\alpha, i'\alpha'\rangle}\!\sum\limits_{s,s'=\pm}\!\!\bigl(\langle s_{\alpha}|s'_{\alpha'}\rangle \tilde{c}^{\dag}_{i\alpha s}\tilde{c}_{i'\alpha' s'}\! + {\rm H. c.}\bigr) \notag\\ 
- J_{\rm H}\sum\limits_{i,\alpha}(\tilde{c}^{\dag}_{i\alpha+}\tilde{c}_{i\alpha+}\! -\! \tilde{c}^{\dag}_{i\alpha-}\tilde{c}_{i\alpha-}).
\label{unperturbed}
\end{eqnarray}
Meanwhile, the perturbed part denotes the electron-magnon interactions, which is composed of higher order terms in $1/S$:
\begin{eqnarray}
\mathcal{H}'\!\! =\! -J_{\rm H}\!\sum\limits_{i,\alpha}\Bigl[\sqrt{\frac{2}{S}}(a_{i\alpha}\tilde{c}^{\dag}_{i\alpha-}\tilde{c}_{i\alpha+}\!\! +\! a^{\dag}_{i\alpha}\tilde{c}^{\dag}_{i\alpha+}\tilde{c}_{i\alpha-}) \notag\\ 
- \frac{1}{S} a^{\dag}_{i\alpha}a_{i\alpha}(\tilde{c}^{\dag}_{i\alpha+}\tilde{c}_{i\alpha+}\!\! -\! \tilde{c}^{\dag}_{i\alpha-}\tilde{c}_{i\alpha-})\Bigr].
\end{eqnarray}
Note that $\mathcal{H}_0$ corresponds to the saddle point Hamiltonian in the variational study~\cite{Akagi2010}, 
and we consider the quantum corrections from $\mathcal{H}'$ systematically below. 

Generalizing the procedure for the ferromagnetic case~\cite{Furukawa1996}, we consider a perturbation expansion in terms of $\mathcal{H}'$.
For the current four-sublattice noncoplanar magnetic order, 
we introduce the magnon Green's function in the matrix form, which includes the anomalous components; 
\begin{eqnarray}
&\hat{\bm D}_{{\mathbf q}}
(\tau) = \begin{bmatrix}
D^{++}_{{\mathbf q}\alpha\alpha'}(\tau) &\!\!\!\!D^{+-}_{{\mathbf q}\alpha\alpha'}(\tau)\\
D^{-+}_{{\mathbf q}\alpha\alpha'}(\tau) &\!\!\!\!D^{--}_{{\mathbf q}\alpha\alpha'}(\tau)
\end{bmatrix} \notag\\
&=\!\!\begin{bmatrix}
-\langle a_{{\mathbf q}\alpha}(\tau)a^{\dag}_{{\mathbf q}\alpha'}(0)\rangle &\!\!\!\!\!-\langle a_{{\mathbf q}\alpha}(\tau)a_{-{\mathbf q}\alpha'}(0)\rangle\\
-\langle a^{\dag}_{-{\mathbf q}\alpha}(\tau)a^{\dag}_{{\mathbf q}\alpha'}(0)\rangle &\!\!\!\!\!-\langle a^{\dag}_{-{\mathbf q}\alpha}(\tau)a_{-{\mathbf q}\alpha'}(0)\rangle
\end{bmatrix},
\end{eqnarray}
where each $D^{\pm\pm}$ is the $4\times 4$ matrix in terms of the sublattice indices $\alpha$ and $\alpha'$; 
${\mathbf q}$ is a wave vector and $\tau$ is an imaginary time. 
The Dyson equation for $\hat{\bm D}_{{\mathbf q}}$ is given by 
$
\hat{{\bm D}}^{-1}_{\mathbf q}(i\omega_{n}) = \hat{{\bm D}}^{(0)-1}_{\mathbf q}(i\omega_{n}) - \hat{{\bm \Sigma}}_{\mathbf q}(i\omega_{n})
$,
where $\hat{{\bm D}}^{(0)}_{\mathbf q}(i\omega_{n}) = (1/i\omega_{n}) \hat{{\bm \tau}}$ is the bare magnon Green's function and 
$\hat{{\bm \Sigma}}_{\mathbf q}(i\omega_n)$ is the magnon self-energy; 
$\hat{{\bm \tau}} = 
\begin{bmatrix} \hat{{\mathbf 1}} & \hat{{\mathbf 0}} \\
\hat{{\mathbf 0}} & -\hat{{\mathbf 1}} \end{bmatrix}$ and 
$\omega_n = 2\pi n/\beta$ ($n$ is an integer and $\beta$ is inverse temperature). 
Here, we consider the Feynman diagrams up to the order of $1/S$. 
Within the lowest order approximation, we can set 
$\hat{{\bm \Sigma}}_{\mathbf q}(i\omega_n) \simeq \hat{{\bm \Sigma}}_{\mathbf q}(0)$; 
the magnon dispersion is obtained from the positive eigenvalues of the Hermitian matrix $\hat{{\bm \Sigma}}_{\mathbf q}(0)$~\cite{Furukawa1996}.

\begin{figure}[t]
\begin{center}
\includegraphics[width=7.0cm]{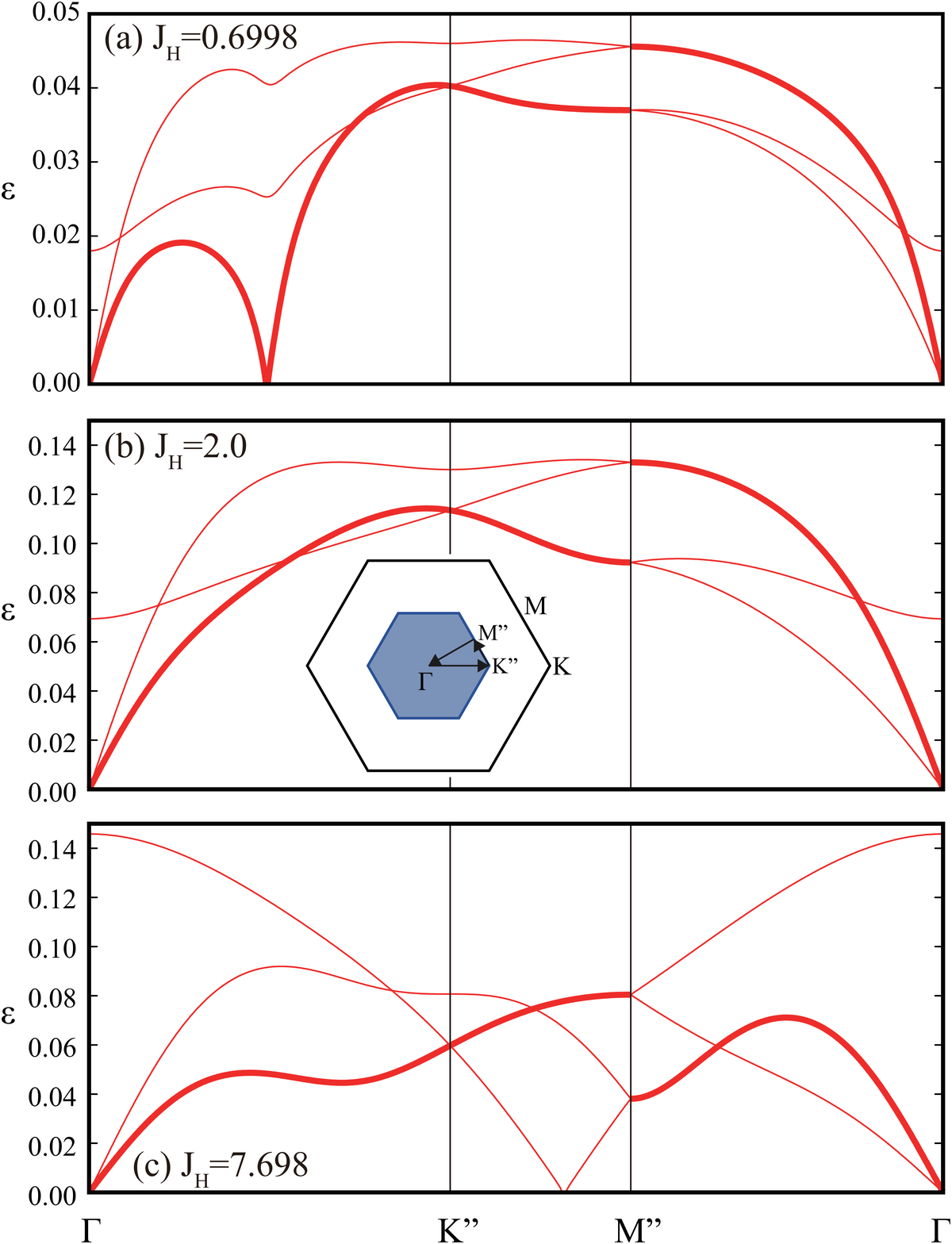}
\end{center}
\caption{(Color online) Magnon dispersions at $n=1/4$ for (a) $J_{\rm H}=0.6998$, (b) $2.0$, and (c) $7.698$.
The thin (thick) curves denote the non-(doubly-)degenerate branches. 
The blue (white) hexagon in the inset of (b) indicates the folded (original) Brillouin zone. 
The magnon dispersions are plotted along the symmetric lines on the folded Brillouin zone. 
}
\label{band_quarter}
\end{figure}

Let us first discuss the results at 1/4 filling, i.e., at $n=\frac{1}{2N}\sum_{i\sigma}\langle c_{i\sigma}^\dagger c_{i\sigma} \rangle=1/4$ ($N$ is the number of sites). 
Within the variational calculation neglecting quantum fluctuations, 
the spin scalar chiral phase at 1/4 filling is stable for $0.7 \lesssim J_{\rm H} \lesssim 4.5$, 
as shown in Fig.~\ref{chiral-order_with-PD}(b). 
Figure~\ref{band_quarter} shows the results of magnon dispersions calculated by the above procedure at 1/4 filling. 
The magnon excitation has three gapless modes and one gapped mode at the $\Gamma$ point.
The three gapless modes are Nambu-Goldstone modes coming from the breaking of $SO(3)$ symmetry. 
As decreasing $J_{\rm H}$, the excitation spectrum changes gradually and shows a softening at or very close to the midpoint on the $\Gamma$-K" line 
for $J_{\rm H} \simeq 0.7$, as shown in Fig.~\ref{band_quarter}(a). 
This signals an instability of the chiral order.
The value of $J_{\rm H}$ for the instability almost coincides with the lower phase boundary in the variational phase diagram, 
where the scalar chiral phase is taken over by a coplanar four-sublattice phase via a narrow phase-separated region [see Fig.~\ref{chiral-order_with-PD}(b)]. 
However, the wave number
of the softening indicates that the instability will occur toward a different ordered state, possibly 
with 48-sublattice sites.
On the other hand, for the upper phase boundary at $J_{\rm H} \simeq 4.5$, 
the magnon excitation does not show any softening; 
the softening takes place at a considerably larger $J_{\rm H} \simeq 7.7$ on the K''-M'' line, as shown in Fig.~\ref{band_quarter}(c). 
Hence, the results indicate that the chiral state obtained by the variational calculation at $n=1/4$ remains stable 
in the entire range of $0.7 \lesssim J_{\rm H} \lesssim 4.5$,
even if quantum fluctuations are taken into account up to the order of $1/S$.

\begin{figure}[t]
\begin{center}
\includegraphics[width=7.0cm]{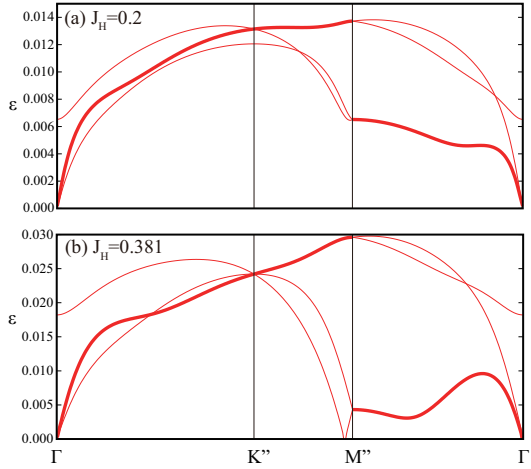}
\end{center}
\caption{(Color online) Magnon dispersions at $n=3/4$ for (a) $J_{\rm H}=0.2$ and (b) $0.381$. 
}
\label{band_three-quarter}
\end{figure}

On the contrary, at 3/4 filling, the spin scalar chiral order found in the variational ground state for $0<J_{\rm H} \lesssim 0.5$ [see Fig.~\ref{chiral-order_with-PD}(b)] 
shows an instability under quantum fluctuations. 
Figure~\ref{band_three-quarter} shows the magnon dispersions at 3/4 filling. 
In this case also, the spectrum is composed of one gapful and three 
gapless modes, and remains stable without showing any softening for $J_{\rm H} \to 0$. 
While increasing $J_{\rm H}$, however, a softening occurs at $J_{\rm H}\simeq0.38$ on the K"-M" line 
at ${\mathbf q}={\mathbf q}^{\ast}\simeq(0.513\pi, 0.266\pi)$. 
The result suggests that the spin scalar chiral phase becomes unstable for $J_{\rm H} \gtrsim 0.38$ and is taken over by an incommensurate magnetic order
with the wave vector ${\mathbf q}^{\ast}$. 
The critical value of $J_{\rm H}$ is substantially smaller than the upper phase boundary of the variational ground state $J_{\rm H} \simeq 0.5$, suggesting that the chiral state is 
reduced by quantum fluctuations.

Our results indicate that the two scalar chiral phases show a contrastive response to quantum spin fluctuations.
The fragility of the 3/4-filling phase may be attributed to the severe competition between different magnetic orders as follows.
Both at 3/4 and 1/4 filling, at the level of the second-order perturbation in terms of $J_{\rm H}/t$,
the chiral phase is energetically degenerate with (2a) two-sublattice collinear stripe and (4f) four-sublattice coplanar 90$^{\circ}$ spiral ordered phases~\cite{Akagi2012}. 
The energy differences between these saddle-point solutions for finite $J_{\rm H}$, however, are one-order of magnitude smaller at 
$3/4$ filling (typically, $\sim 10^{-3}t$) than at 1/4 filling ($\sim 10^{-2}t$). 
This severe competition makes the 3/4-filling phase more fragile against quantum fluctuations compared to the 1/4-filling one. 
Similar contrastive response was also found for the modulation of the lattice structure~\cite{Akagi2013}. 
These results imply that the chiral Chern insulating state at $n=1/4$, 
which extends in a much wider $J_{\rm H}$ region than the 3/4-filling one, is robust against perturbations.

Next, we discuss the effect of zero-point oscillation to the length of ordered moment in the spin scalar chiral phase.
Due to the equivalence of sublattices, the ordered moment is uniformly reduced by 
$\Delta S = S - \langle \tilde{S}^z \rangle =\frac{1}{N}\sum_{i,\alpha} \langle a^{\dag}_{i\alpha} a_{i\alpha} \rangle 
= \frac{1}{N}\sum_{{\mathbf q},\alpha} \langle a^{\dag}_{{\mathbf q}\alpha}a_{{\mathbf q}\alpha}\rangle$,
which is calculated by
\begin{eqnarray}
\Delta S=
\frac{1}{N}\sum\limits_{{\mathbf q},\alpha} \sum_{\eta}
\frac{|T_{{\mathbf q}\alpha\eta}|^2}{e^{\beta \varepsilon_{{\mathbf q}\eta}}-1} {\rm sign}({\varepsilon_{{\mathbf q}\eta}}). 
\end{eqnarray}
Here, ${\rm sign}(x)=1 (-1)$ for $x>0$ ($x<0$), 
$\varepsilon_{{\mathbf q}\eta}$ is the eigenvalue of magnon self-energy, and $T_{{\mathbf q}\alpha\eta}$ is the matrix element of the para-unitary matrix $\hat{{\bm T}}_{\mathbf q}$ 
which diagonalizes the magnon self-energy, i.e., 
$\hat{{\bm T}}^{-1}_{\mathbf q} \hat{{\bm \tau}} \hat{{\bm \Sigma}}_{\mathbf q}(0) \hat{{\bm T}}_{\mathbf q} =\hat{{\bm \tau}} \hat{{\bm \epsilon}}_{\mathbf q}$ 
under the condition $\hat{{\bm T}}^{\dag}_{\mathbf q}\hat{{\bm \tau}} \hat{{\bm T}}_{\mathbf q} = \hat{{\bm \tau}}$~\cite{Colpa1978}.

\begin{figure}[t]
\begin{center}
\includegraphics[width=7.0cm]{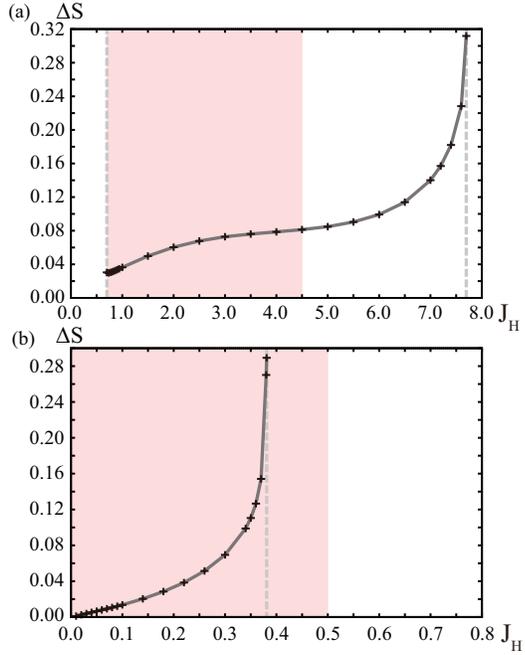}
\end{center}
\caption{(Color online) The moment reduction of the chiral state at zero temperature for (a) $n=1/4$  and (b) $n=3/4$. 
The shaded areas indicate the range of $J_{\rm H}$ in which the variational ground state shows the chiral order [see Fig.~\ref{chiral-order_with-PD}(b)]. 
The vertical dashed lines denote the values of $J_{\rm H}$ at which the magnon excitations show the softenings.	
}
\label{moment-reduction}
\end{figure}

Figure~\ref{moment-reduction} shows the reduction of ordered moments, $\Delta S$, at 1/4 and 3/4 filling.
The shaded areas indicate the range of $J_{\rm H}$ for the chiral ordered phases in the variational calculation
[see Fig.~\ref{chiral-order_with-PD}(b)]. 
As shown in the figures, $\Delta S$ remains substantially small except in the vicinity of the softening~\cite{comment}. 
In particular, at 1/4 filling [Fig.~\ref{moment-reduction}(a)], $\Delta S$ is less than $0.082$ 
for all the values of $J_{\rm H}$ in the shaded chiral region. 
The results indicate that the moment reduction in the chiral phases is small and the localized moments are well approximated by classical spins.

Let us compare the value of $\Delta S$ with that in the absence of the chiral order. 
In the case of the $120^{\circ }$ N{\'e}el order at half filling~\cite{Akagi2013_10}, which is noncollinear but coplanar, 
$\Delta S$ is relatively large, not substantially suppressed from the value for the antiferromagnetic Heisenberg model, 
$\Delta S \simeq  0.261$~\cite{Jolicoeur1999}. 
Our result for the chiral ordered phase in Fig.~\ref{moment-reduction}(a) is almost three times smaller than the coplanar case.
This large difference implies that the additional $Z_2$ symmetry breaking related to the chiral ordering plays a role in the robustness against quantum fluctuations. 
Note that similar robustness was discussed for thermal fluctuations~\cite{Kato2010}.

In Fig.~\ref{moment-reduction}(b), it is 
interesting to note that, at 3/4 filling, $\Delta S$ approaches zero as $J_{\rm H} \to 0$.
This suggests that quantum spin fluctuations are irrelevant in the weak coupling limit. 
The detailed analysis will be reported elsewhere. 

To summarize, we have investigated the effect of quantum spin fluctuations on 
the spin scalar chiral order in the Chern insulating phases, which were found in the Kondo lattice model on a triangular lattice. 
Within the linear spin wave analysis, we have found that the stability of the chiral phases is largely dependent on the electron filling; 
that is, the 3/4-filling chiral state is destabilized by quantum fluctuations, whereas the 1/4-filling one remains stable. 
We have also found that the reduction of the ordered moment is considerably smaller than the cases in the absence of itinerant electrons. 
The results indicate that the Chern insulator with spin chiral ordering at 1/4 filling is robust under quantum fluctuations. 

The robustness will be beneficial for the realization of the chiral order in real materials. In fact,
there are several candidates of itinerant magnets that possess triangular layers, 
such as delafossite oxides and their relatives~\cite{Foo2004,Okuda2005,Okuda2008,Takatsu2009,Takatsu2010} and 
materials with a hexagonal CeCo$_4$B- or CaCu$_5$-type structure~\cite{Mentink1994,Salamakha2013}.  
Thin films of face-centered-cubic magnets, such as double perovskites and rocksalt oxides, are also worth investigating. 
Our results will stimulate the exploration of the chiral phase in these materials and the exotic phenomena inherent in this phase, such as 
the quantized anomalous Hall response~\cite{Martin2008,Akagi2010} and fractional excitations~\cite{Venderbos2012,Rahmani2013}.

\acknowledgements
{
The authors acknowledge helpful discussions with Cristian D. Batista, Takahiro Misawa, Joji Nasu, Nic Shannon, and Youhei Yamaji. 
Y.A. is supported by Grant-in-Aid for JSPS Fellows.
This work was supported by Grants-in-Aid for Scientific Research (Grants Nos. 24340076 and 24740221), 
the Strategic Programs for Innovative Research (SPIRE),
MEXT, and the Computational Materials Science Initiative (CMSI), Japan.
}


\begin{thebibliography}{30}
\bibitem{comment} $\Delta S$ does not diverge even at the softening in the two-dimensional case. 
We note that there is a small enhancement near $J_{\rm H} \simeq 0.7$ in the case of 1/4 filling.
\end{thebibliography}
\end{document}